\documentclass[%
 reprint,
superscriptaddress,
 amsmath,amssymb,
 aps,
]{revtex4-2}
\usepackage{physics}
\usepackage{bbold}
\usepackage{amsmath}
\usepackage{latexsym}
\usepackage{graphicx}
\usepackage{amsthm}
\usepackage{hyperref}
\usepackage{lipsum}
\usepackage[english]{babel}
\usepackage{slashed}
\usepackage[compat=1.1.0]{tikz-feynman}
\usepackage{appendix}
\usepackage{subfig}
\usepackage{multirow}
\usepackage{mathtools}
\usepackage{subfig}
\usepackage{pgfplots}
\pgfplotsset{compat=1.15}
\usepackage{mathrsfs}
\usetikzlibrary{arrows}
\definecolor{ffffff}{rgb}{1,1,1}
\definecolor{qqqqff}{rgb}{0,0,1}
\definecolor{qqwuqq}{rgb}{0,0.39215686274509803,0}
\definecolor{zzttqq}{rgb}{0.6,0.2,0}
\definecolor{ttqqqq}{rgb}{0.2,0,0}
\definecolor{qqttqq}{rgb}{0,0.2,0}
\definecolor{qqccqq}{rgb}{0,0.8,0}
\usepackage{tikz-cd}
\usepackage{amsmath}
\usepackage{amsbsy}

\theoremstyle{plain}

\tikzset{
pattern size/.store in=\mcSize, 
pattern size = 5pt,
pattern thickness/.store in=\mcThickness, 
pattern thickness = 0.3pt,
pattern radius/.store in=\mcRadius, 
pattern radius = 1pt}
\makeatletter
\pgfutil@ifundefined{pgf@pattern@name@_b8864pxwu}{
\pgfdeclarepatternformonly[\mcThickness,\mcSize]{_b8864pxwu}
{\pgfqpoint{0pt}{0pt}}
{\pgfpoint{\mcSize+\mcThickness}{\mcSize+\mcThickness}}
{\pgfpoint{\mcSize}{\mcSize}}
{
\pgfsetcolor{\tikz@pattern@color}
\pgfsetlinewidth{\mcThickness}
\pgfpathmoveto{\pgfqpoint{0pt}{0pt}}
\pgfpathlineto{\pgfpoint{\mcSize+\mcThickness}{\mcSize+\mcThickness}}
\pgfusepath{stroke}
}}
\makeatother

\tikzset{
pattern size/.store in=\mcSize, 
pattern size = 5pt,
pattern thickness/.store in=\mcThickness, 
pattern thickness = 0.3pt,
pattern radius/.store in=\mcRadius, 
pattern radius = 1pt}
\makeatletter
\pgfutil@ifundefined{pgf@pattern@name@_f6j18z8qd}{
\pgfdeclarepatternformonly[\mcThickness,\mcSize]{_f6j18z8qd}
{\pgfqpoint{0pt}{0pt}}
{\pgfpoint{\mcSize+\mcThickness}{\mcSize+\mcThickness}}
{\pgfpoint{\mcSize}{\mcSize}}
{
\pgfsetcolor{\tikz@pattern@color}
\pgfsetlinewidth{\mcThickness}
\pgfpathmoveto{\pgfqpoint{0pt}{0pt}}
\pgfpathlineto{\pgfpoint{\mcSize+\mcThickness}{\mcSize+\mcThickness}}
\pgfusepath{stroke}
}}
\makeatother
\tikzset{every picture/.style={line width=0.75pt}} 

\begin{document}

\title{Time and distance constraints for mass and charge interferometry}
\author{Adrian Kent}
\affiliation{Centre for Quantum Information and Foundations, DAMTP, Centre for Mathematical Sciences, University of Cambridge, Wilberforce Road, Cambridge CB3 0WA, UK}
\affiliation{Perimeter Institute for Theoretical
	Physics, 31 Caroline Street North, Waterloo, ON N2L 2Y5, Canada.}
\date{\today}

\begin{abstract}
We reanalyse and extend constraints on mass and charge interferometry
identified by Mari et al. (2016). 
We show that their constraint on the time required for coherent interference
can be extended by a factor of two.   We extend their analysis to consider
experiments in which one interferometer measures gravitational or electric fields
generated by another.   We note that these analyses imply
a maximum separation between a mass or charge interferometer and a decohering
gravitational or electric field measurement that can be carried out without backreaction.   
\end{abstract}
\maketitle

\section{Introduction}

In a very interesting paper, Mari et al. \cite{mari2016experiments} (MPG) consider the apparently
paradoxical fact that mass and charge interferometry is possible, even
though the particles involved have measurable 
long-range gravitational and electrostatic fields.
One might think that the fields are thus entangled with the
particles, and effectively decohere them, preventing interference 
from being observed.
This sense of paradox is sharpened by relativistic causality.
If we model the particles as classical sources, the 
field disturbances associated with different interferometer paths
propagate at light speed into the far distance and future, so that
recombining the particles apparently cannot erase the records in
the fields completely, even if it does so in their vicinity.
Moreover, whether or not the fields are ever measured, it seems
that they could be at any point during or after the experiment,
giving a macroscopically amplified outcome that determines the path.   

Focusing on this last point, they propose the following resolution.  
A field measurement requires quantum matter to be coupled
to the local field, and its state to be observed.   This cannot
be done instantaneously, and in fact, in the model of measurement
they offer, it requires a minimum time for a given precision.  
Let the time required for a given interferometry experiment in
region $A$ be $T_A$, and let $T_B$ be the time required to
measure precisely enough a point $B$ to distinguish the 
interfering paths.  
If 
\begin{equation}\label{maribound}
 T_A + T_B > {{ d(A,B) } \over {c}}   
\end{equation}
then, they argue, there
is no violation of relativistic causality.    
This allows them to calculate lower bounds $T_A^{\rm min}$ on the time $T_A$
(as a function of the mass $m_A$ or charge $q_A$ and path
separation $d$)
for a successful interference experiment to be possible.
Significantly faster interference experiments will, they
argue, decohere through radiation, whether or not the field
is measured elsewhere.

Here we refine their insights.   
First, we note that their analysis of causal relations between
the experiments at $A$ and $B$ can be extended, giving essentially
an extra factor of $2$ in (\ref{maribound}) and a larger factor
in their bound on $T_A^{\rm min}$. 
Second, we consider measurements in which $B$ uses an interferometer
(measuring phase) rather than a coupled particle (measuring displacement), 
and derive bounds for this case. 
Third, we note that all these bounds imply a maximum distance $d(A,B)$
for the measurement at $B$ to be able to decohere the interferometer at $A$
with no back-reaction of the objects involved in $B$'s measurement affecting
$A$'s particle.   
In other words, although the classical field disturbance propagates 
arbitrarily far in distance and time, beyond a certain point a 
sufficiently precise measurement takes long enough
that the potentials generated during this measurement process 
in turn affect the interferometer.   

\section{Varying the causal relations in Mari et al.'s experimental set-up}

We begin, following MPG's discussion and notation, by considering a 
thought experiment
involving two separated parties, Alice and Bob, who are both stationary
in the same inertial frame.   We focus on mass interferometry:
the discussion of charge interferometry is analogous. 
For now we work in one-dimensional space.  
Alice carries out a mass interferometry experiment that involves 
causing a mass $m_A$ to follow two paths, separated by $d$ during
the relevant part of the experiment.  We write the state of 
her particle as
\begin{equation}\label{super}
    \ket{\psi} = {1 \over \sqrt{2}}  ( \ket{L} + \ket{R} ) \,  .
\end{equation}
Here $\braket{x}{L} = \phi (x) $, $\braket{x}{R} = \phi (x-d) $,
where $\phi$ is sharply peaked about $x=0$, $\Delta_x (\phi ) \ll d$, 
and we can take $\phi$ effectively constant throughout.  
We take $d(A,B) = R$.   For the approximations below we
require 
\begin{equation}\label{rbig}
R \gg d \, .
\end{equation}
We neglect all gravitational fields other than those sourced by $m_A$.

A crucial assumption in the following analysis is that Alice's
interferometry does not irreversibly generate entanglement between
her particle state and the gravitational field state.  In particular,
the particle should not radiate in a way that allows the path to 
be inferred from the radiation.   For the
case of electromagnetism, MPG offer arguments that Alice can ensure this, 
by arranging suitably smooth interferometry paths with suitably small acceleration,
if $T_A$ is larger than the maximum value derivable from (\ref{maribound}).
We plan to look in more detail at those arguments, both for electromagnetism and 
gravity, elsewhere.   It is enough here to note that, if and when those
arguments fail, Alice's interferometry will fail, since the paths are 
decohered by the field.   

Bob prepares a test mass $m_B$ in the ground state of a narrow harmonic trap.
He may choose to measure the local gravitational field at any time by 
removing the trap, allowing his particle state to evolve in the local
gravitational potential.   If he does this then, 
after a time $T_B$, which depends on $d,R$ and $m_A$,
it evolves to one of two near-orthogonal states $\ket{\psi_L}$ or $ \ket{\psi_R}$, depending
on whether Alice's particle is in state $\ket{L}$ or $\ket{R}$.
He completes his measurement by observing whether his 
particle is in state $\ket{\psi_L}$ or $ \ket{\psi_R}$.

Consider the following timings, which differ from those discussed by MPG.
Alice creates the superposition (\ref{super}) 
by time $t= {- R \over c}$ and maintains it until time $t={- R \over c } + T_B$.
Bob decides at time $t=0$ whether or not to remove the trap and initiate a
measurement.   If he does, it is complete at time $T_B$.    Note that 
relativistic causality implies that Bob's measurement at time $t$ is 
effectively of Alice's particle state at time $t- { R \over c }$, since 
the effect of any change of the particle state on the gravitational
field takes time ${R \over c}$ to propagate from Alice to Bob. 
Alice completes her interference experiment by effectively recombining her 
particle's paths.  (This may involve operations of her choice: 
for example, she may couple the paths to orthogonal 
spin states of a spin ${1 \over 2}$ system and measure the final
spin.)   This takes time $T_A$, so is completed at time 
$t={- R \over c } + T_A + T_B$.   

Now, as MPG note, in any configuration, relativistic causality requires 
that Alice should not
be able to learn whether or not Bob measured until time ${R \over c}$
after his decision.   If Bob measures, effectively measuring 
Alice's particle in state $\ket{L}$ or $\ket{R}$, then from her
perspective her particle will be in the mixture
\begin{equation}\label{mix}
 \rho = {1 \over 2}  ( \ket{L} \bra{L} + \ket{R} \bra{R}) \,  , 
\end{equation}
which is probabilistically distinguishable from (\ref{super}).  
If he does not, the particle remains in state (\ref{super}).
Since Bob's decision is made at $t=0$, this gives that
\begin{equation}
{- R \over c } + T_A + T_B >  { R \over c}  
\end{equation}
and so
\begin{equation}\label{newbound}
   T_A + T_B >  {2 R \over c} \, ,   
\end{equation}
improving on (\ref{maribound}) by a factor of $2$.

Now MPG note that, when (\ref{rbig}) holds, the differential force 
between the paths at B is
\begin{equation}
\Delta F \approx {{ G m_A m_B d} \over {R^3}} \, , 
\end{equation}
which produces a position shift 
\begin{equation}
\delta_x \approx {{ \Delta F T_B^2 } \over {2 m_B}} \approx {{ Gm_A T_B^2 d} \over {2 R^3}}
\end{equation}
after time $T_B$.

Assuming that Bob's trap cannot confine the particle with position uncertainty
\begin{equation}
    \Delta X > l_P = ( {{ \hbar G } \over { c^3 } } )^{1 \over 2} \, , 
\end{equation}
this means $T_B$ must satisfy 
\begin{equation}\label{tbone}
    {{ m_A d c^2 T_B^2 }  \over { 2 m_P R^3}}  \gtrsim 1 \, . 
\end{equation}
Since (\ref{newbound}) implies no constraint on $T_A$ when 
$T_B \geq {2 R \over c}$, we are interested in the 
case $T_B = \eta {2 R \over c}$ for $0 < \eta < 1$.
We are also interested in minimizing $T_B$ as far as physically possible,
so we take 
\begin{equation}\label{tbtwo}
    {{ m_A d c^2 T_B^2 }  \over { 2 m_P R^3}}  = 1 \, . 
\end{equation}
From (\ref{tbone},\ref{tbtwo},\ref{newbound}) we get
\begin{equation}\label{tbthree}
T_B = {{ 8 \eta^3 m_A d} \over {2 m_P c}}  \, 
\end{equation}
and
\begin{equation}
   T_A > 4 ( \eta^2 - \eta^3 )  {{m_A d} \over { m_P c}}   \, ,    
\end{equation}
Maximizing over $\eta$ gives 
\begin{equation}
   T_A >  { 16 \over 27 }  {{m_A d} \over { m_P c}}   \, ,    
\end{equation}
a factor of $8$ larger than MPG's Eqn. (20), although of course
of the same order in fundamental physical parameters.   

\section{Bound on the separation of a non-disturbing measurement}

For there to be no gravitational back-reaction from $B$'s system
on $A$'s particle in the course of $B$'s measurement, we require
\begin{equation}
T_B < {R \over c} \, . 
\end{equation}
Combining this with (\ref{tbone}) gives
\begin{equation}
R \leq {{m_A d} \over {2 m_P}} \, .
\end{equation}
Since $R \gg d$, we need $m_A \gg m_P$ for any 
back-reaction free measurement to be possible.    

\section{Measuring the gravitational field by interferometry}

Now suppose that $B$ is able to set up an idealized mass interferometer to measure
the gravitational field generated by $A$'s particle. 
At $t=0$, $B$ either leaves his mass $m_B$ in an unsuperposed position state
$\ket{L}_B$ at distance $R$ from $A$, or splits the beam into a superposition ${1 \over \sqrt{2}} (\ket{L}_B + \ket{R}_B)$.    We suppose, since this is the most favourable possible case for fast measurement, that the $\ket{R}_B$ beam moves away from $A$ fast
enough that the phase difference arising from its gravitational interaction
with $A$'s particle's two paths during $t>0$ is negligible.  

Note that, while this is justifiable for deriving the bound below, it may not
be possible for a physically realisable interferometry measurement, since
the required beam acceleration may generate radiation. 
Conservatively neglecting this, 
$B$'s state then evolves into one of two states of the form
\begin{eqnarray}
\ket{\psi_L}_B &=& {1 \over \sqrt{2}} ( \exp( i \phi_L ) \ket{L}_B + \ket{R}_B ) \, , \\
\qquad \ket{\psi_R}_B &=& {1 \over \sqrt{2}} ( \exp( i \phi_L ) \exp (i \Delta \phi_L ) \ket{L}_B + \ket{R}_B ) \nonumber \, , 
\end{eqnarray}
where
\begin{equation}
\Delta \phi_L = {{ G m_A m_B T_B} \over {\hbar}} ( {1 \over R} - { 1 \over {R+d }} )
\approx {{ G m_A m_B T_B d} \over {\hbar R^2}} \, . 
\end{equation}

These are approximately orthogonal when $\Delta \phi_L \approx \pi$.
Assuming that $B$ can recombine
the beams in relatively negligible time, which is the most
favourable possibility for fast measurement (though again may be unphysical), this 
would give
\begin{equation}
T_B \approx {{\hbar R^2 \pi} \over { G m_A m_B d}}  \, . 
\end{equation}

For no back-reaction affecting the phases of $A$'s particles we require
$T_B < {R \over c}$, which gives 
\begin{equation}
R < {{m_A m_B d } \over {\pi m_P^2}} \, .
\end{equation}
As we also require (\ref{rbig}), we need
\begin{equation}
m_A m_B \gg m_P^2
\end{equation}
and so we must have $m_A \gg m_P $ or $m_B \gg m_P$
for the possibility of a back-reaction free measurement.

\section{Discussion}

We have explored MPG's insights further, improving their quantitative results, 
while obtaining results consistent with the fundamental physical parameter
dependence of theirs.   Each experiment discussed gives interesting bounds, 
which are consistent both with relativistic causality and with the intuition
that which-path measurements via gravitational fields are possible in 
principle.   

This is not a proof that every experiment that could possibly
be considered produces the same conclusions.   Indeed, the bounds
could be modestly further improved if one considered two experiments by
Bobs equidistant from and on opposite sides of Alice's interferometer, 
and perhaps further improved by considering other configurations.      
One might wonder whether more dramatic improvements could be obtained
by, for example, considering a dense array of $N$ Bobs in a 
spherical shell segment with small solid angle separated from
Alice by a distance in a small interval around $R$.  
However, while one might naively hope this gives a $\sqrt{N}$ speedup, 
one has to allow for the interactions among the Bobs' measuring devices,
which themselves become entangled.    
It would be interesting to explore this and other possibilities in detail,
in the hope of proving general and tight bounds.    

Our discussion reinforces the conclusion that back-reaction free measurements
of gravitational fields from a mass interferometer require at least one 
mass interferometer (either the measured one or one used as a measuring 
device) to involve a mass $m \gg m_P$, which unfortunately appears far
beyond foreseeable technology.   

\bigskip

\section{Acknowledgements}
I acknowledge financial support from the
UK Quantum Communications Hub grant no. 
EP/T001011/1. 
This work was supported in part by Perimeter Institute for
Theoretical Physics. Research at Perimeter Institute is supported by
the Government of Canada through the Department of Innovation, Science
and Economic Development and by the Province
of Ontario through the Ministry of Research, Innovation and Science.
I am very grateful to Carlo Rovelli for drawing my attention to 
Ref. \cite{mari2016experiments} and for many invaluable comments
and insights.   

\bigskip

\bibliographystyle{unsrt}
\bibliography{library,library2}

\end{document}